\documentclass[
a4paper,
aps,
prd,
twocolumn,
tightenlines,
preprintnumbers,
nofootinbib,
showkeys,
superscriptaddress
]{revtex4-1}

\usepackage{XCharter}
\usepackage[T1]{fontenc}

\usepackage{fullpage}
\usepackage{amsfonts}
\usepackage{amsmath}
\usepackage{slashed}
\usepackage{amssymb}
\usepackage{graphicx}
\usepackage{epic}
\usepackage{eepic}
\usepackage{epsfig}
\usepackage{latexsym}
\usepackage[dvipsnames]{xcolor}
\usepackage{float}
\usepackage{multirow}
\usepackage{hyperref}
\usepackage{enumitem}
\hypersetup{colorlinks=true,citecolor=red,linkcolor=NavyBlue,urlcolor=NavyBlue}
\usepackage[caption=false]{subfig}
 
\usepackage{natbib}
\usepackage{relsize}
\usepackage[left=1.9cm,right=1.9cm,top=2.0cm,bottom=2.5cm]{geometry}
\usepackage{shorthand}

\begin{document}
\relscale{0.95}

\title{Hunting scalar leptoquarks with boosted tops and light leptons}

\author{Kushagra Chandak}
\email{kushagra.chandak@research.iiit.ac.in}
\affiliation{Center for Computational Natural Sciences and Bioinformatics, International Institute of Information Technology, Hyderabad 500 032, India}

\author{Tanumoy Mandal}
\email{tanumoy.mandal@physics.uu.se}
\affiliation{Department of Physics and Astronomy, Uppsala University, Box 516, SE-751 20 Uppsala, Sweden}

\author{Subhadip Mitra}
\email{subhadip.mitra@iiit.ac.in}
\affiliation{Center for Computational Natural Sciences and Bioinformatics, International Institute of Information Technology, Hyderabad 500 032, India}

\date{\today}

\begin{abstract}
\noindent
The LHC search strategies for leptoquarks that couple dominantly to a top quark are different than for the ones that couple mostly to the
light quarks. We consider charge $1/3$ ($\phi_1$) and $5/3$ ($\phi_5$) scalar leptoquarks that can
decay to a top quark and a charged lepton ($t\ell$) giving rise to a resonance system of a boosted top and a 
high-$p_{\rm T}$ lepton. We introduce simple phenomenological models suitable for bottom-up studies and explicitly map them to all possible scalar leptoquark models within
the Buchm\"{u}ller-R\"{u}ckl-Wyler classifications that can have the desired decays. We study pair and single productions of these leptoquarks.
Contrary to the common perception, we find that the single production of top-philic leptoquarks $\phi = \{\phi_1,
\phi_5\}$ in association with a lepton and jets could be significant for order one $\phi t\ell$ coupling in certain scenarios. 
We propose a strategy of selecting events with at least one hadronic-top and two high-$p_{\rm T}$ same flavour opposite sign leptons. This captures events from
both pair and single productions. Our strategy can significantly enhance the LHC discovery potential
especially in the high-mass region where single productions become more prominent.
Our estimation shows that a scalar leptoquark as heavy as $\sim1.7$ TeV can be discovered at the $14$ TeV LHC with 
3 ab$^{-1}$ of integrated luminosity in the $t\ell\ell+X$ channel for $100\%$ branching ratio in the $\phi\to t\ell $ decay mode. 
However, in some scenarios, the discovery reach can 
increase beyond $2$ TeV even though the branching ratio comes down to about $50\%$.
\end{abstract}

%\keywords{Leptoquarks, Top-tagging, HL-LHC}

\maketitle

%%%%%%%%%%%%%%%%%%%%%%%%%%%%%%%%%%%%%%%%%%%%%%%%%%%%%%%%%%%%%%%%%%%%%%%%%%%%%%%%%%%%%%%%%%%%%%%%%%%%%%%%%%%%%%%%%

\section{Introduction}
\label{sec:intro}

\noindent
So far, the predictions of the Standard Model (SM) have been verified to a
remarkable degree of accuracy. But some persistent deviations in rare $B$-meson decays
observed in several independent experiments hint towards new physics. 
In particular, a significant excess in the $R_{D^{(*)}}$ observables was first reported by the BaBar
collaboration in 2012~\cite{Lees:2012xj,Lees:2013uzd}. Later, this excess was also seen in the 
LHCb~\cite{Aaij:2015yra,Aaij:2017uff,Aaij:2017deq} and Belle~\cite{Hirose:2016wfn,Hirose:2017dxl,Abdesselam:2019dgh}
measurements (though its significance reduced). The current combined deviation in the $R_{D}$ and $R_{D^{*}}$ observables, as computed by the
HFLAV group~\cite{Amhis:2016xyh}, is still about $3.1\sg$ away from the SM 
prediction~\cite{Bigi:2016mdz,Bernlochner:2017jka,Bigi:2017jbd,Jaiswal:2017rve}. 
In the $R_{K^{(*)}}$ observables, a deviation of about $2.5\sg$ from the corresponding SM 
predictions~\cite{Hiller:2003js,Bordone:2016gaq} have been observed by the LHCb 
collaboration~\cite{Aaij:2014ora,Aaij:2014pli,Aaij:2015oid,Aaij:2017vbb,Aaij:2019wad}. Altogether, 
these deviations indicate towards lepton universality violation and suggest that the
underlying new physics, if that really is the origin of these anomalies, has strong affinity
towards the third generation SM fermions.

A popular explanation of the rare $B$-decay anomalies is the existence of TeV-scale scalar
leptoquarks (LQ or $\ell_q$) that has large couplings to the third generation quarks.
LQs  appear in different scenarios like 
Pati-Salam  models \cite{Pati:1974yy}, $\mathrm{SU}(5)$ grand unified theories \cite{Georgi:1974sy}, 
the models with quark lepton compositeness \cite{Schrempp:1984nj},  $R$-parity violating supersymmetric models 
\cite{Barbier:2004ez} or coloured Zee-Babu model \cite{Kohda:2012sr} etc. Their phenomenology has also 
been studied in great detail (see, for example, Refs.~\cite{Gripaios:2010hv,Arnold:2013cva,Bandyopadhyay:2018syt,Vignaroli:2018lpq,Mandal:2018kau,Roy:2018nwc,Alves:2018krf,Aydemir:2019ynb} 
for some phenomenological studies).

The LHC is actively looking for the signatures of scalar LQs that couple with third generation fermions for some time and has put direct 
bounds on them. Among the various possible signatures, the 
$pp\to\ell_{q}\ell_{q}\to tt\tau\tau$ mode is already extensively searched for by the ATLAS and the CMS collaborations. Assuming $100$\% branching ratio (BR) in the $\ell_{q}\to t \tau$ decay mode, the latest scalar LQ pair production 
search at the CMS detector has excluded masses below $900$ GeV \cite{Sirunyan:2018nkj}. CMS has also put bounds 
on scalar LQs that decay to a $b$-quark and a neutrino at about $1.1$ TeV assuming $100\%$ BR in this decay mode~\cite{Sirunyan:2018kzh}.
Similar limits are also available from the ATLAS searches~\cite{Aaboud:2019jcc,Aaboud:2019bye}.

In this paper, we consider scalar LQs with a non-standard decay to a top quark and a light charged lepton ($\ell=\{e,\m\}$). In light of the observed $B$-decay anomalies, such non-standard decay modes have started getting some attention. For example, the CMS collaboration has recently published their first analysis of LQ pair production searches in the $tt\mu\mu$ channel~\cite{Sirunyan:2018ruf}. 
They have also done a prospect study for this channel at the HL-LHC based on the $13$ TeV data collected in 2016~\cite{CMS:2018yke}. Generally,  it is possible to
have LQs with large cross-generational couplings i.e. a LQ that couples to quarks and leptons
of different generations~\cite{Diaz:2017lit,Schmaltz:2018nls}. However, large cross-generational couplings would introduce 
flavour changing neutral currents which are strongly constrained from precision experiments except for the cases where LQs couple with third generation quarks. 
For the light  lepton we consider either an electron or a muon but not both at the same time.
This is because, the scenarios with comparable couplings of a LQ to leptons of different generations simultaneously 
(and hence comparable BRs to those modes) would be constrained by the lepton number/flavour violation experiments. 
With this, pair production of such LQs would have either of the two possible signatures viz. $tt\mu\mu$ and $ttee$.

In this paper, we look beyond the pair production process of scalar LQs and  consider their single productions also. 
The motivation for this is twofold. First, as the LQ mass increases, the pair production cross section falls of faster than the single 
production cross sections due to the extra phase space suppression it receives. Second, the recent $B$-decay anomalies indicate
towards the presence of large cross-generational couplings of LQs -- a necessary condition to search for the single production 
processes. However, the common perception is that LQs that couple with third generation quarks exclusively would have tiny 
single production cross sections for perturbative new couplings because of the small $b$-quark parton density function (PDF) 
($t$-PDF is absent). Here, we implement a search strategy \cite{Mandal:2012rx,Mandal:2015vfa,Mandal:2016csb} by 
combining events of pair and single productions of scalar LQs  in the signal. 
We use a publicly available dedicated top-tagger to tag hadronically decaying boosted tops in the final states 
and estimate the LHC discovery potential of LQs in the  $t\ell\ell +X$ mode.
Contrary to the  common perception, we find that if the unknown couplings controlling the single production processes are not 
very small but perturbative (i.e. order one), such a strategy can enhance the discovery prospect of LQs at the LHC 
significantly.

The rest of the paper is organized as follows. In Section~\ref{sec:model}, we introduce the 
leptoquark models. In Section~\ref{sec:pheno} we discuss the LHC phenomenology and  our search strategy 
and present our results in Section~\ref{sec:dispot}. Finally, we summarise and conclude in Section~\ref{sec:End}.

\begin{figure*}
\captionsetup[subfigure]{labelformat=empty}
\subfloat[(a)]{\includegraphics[height=3cm,width=4cm]{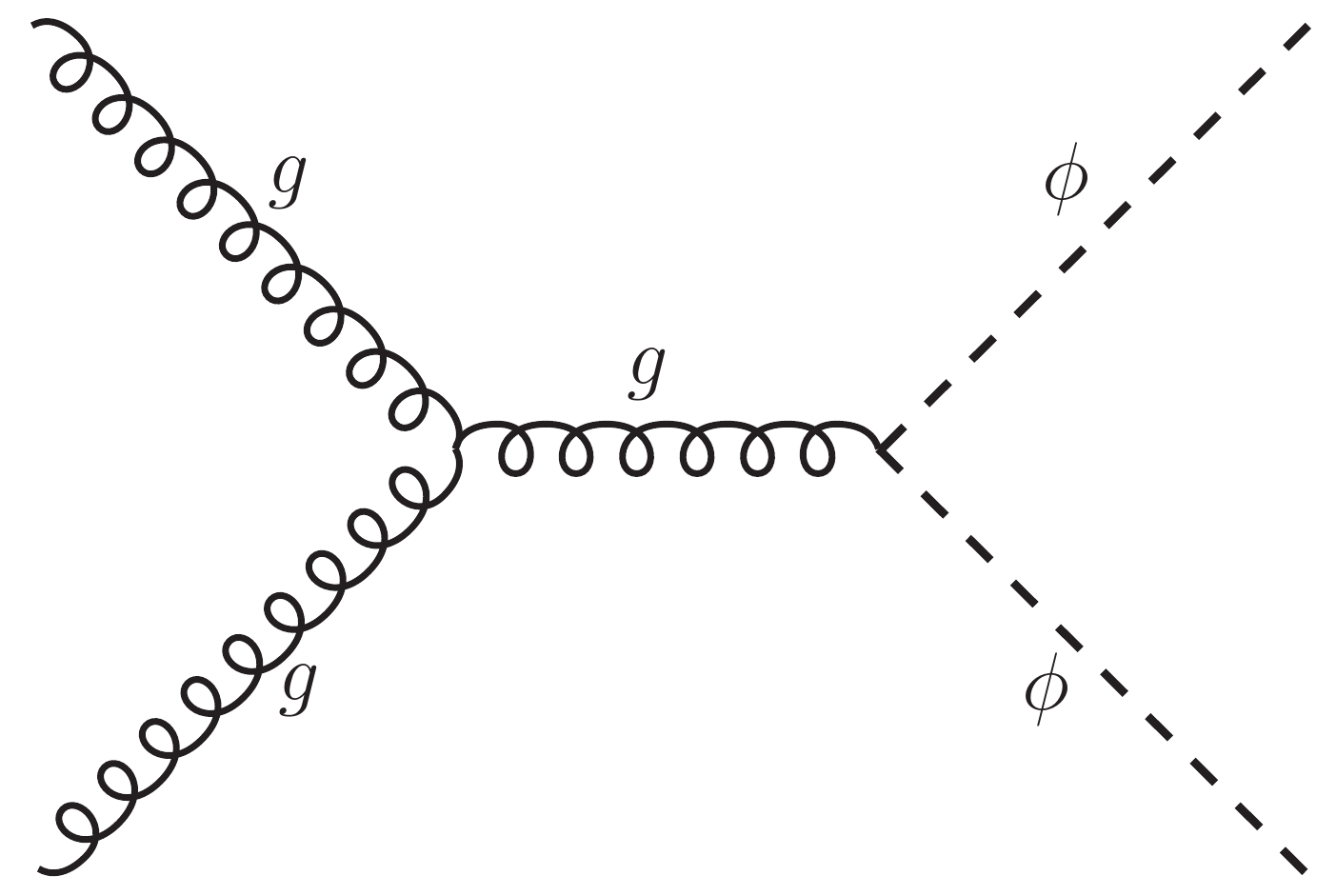}\label{fig:feyna}}\hfill
\subfloat[(b)]{\includegraphics[height=3cm,width=4cm]{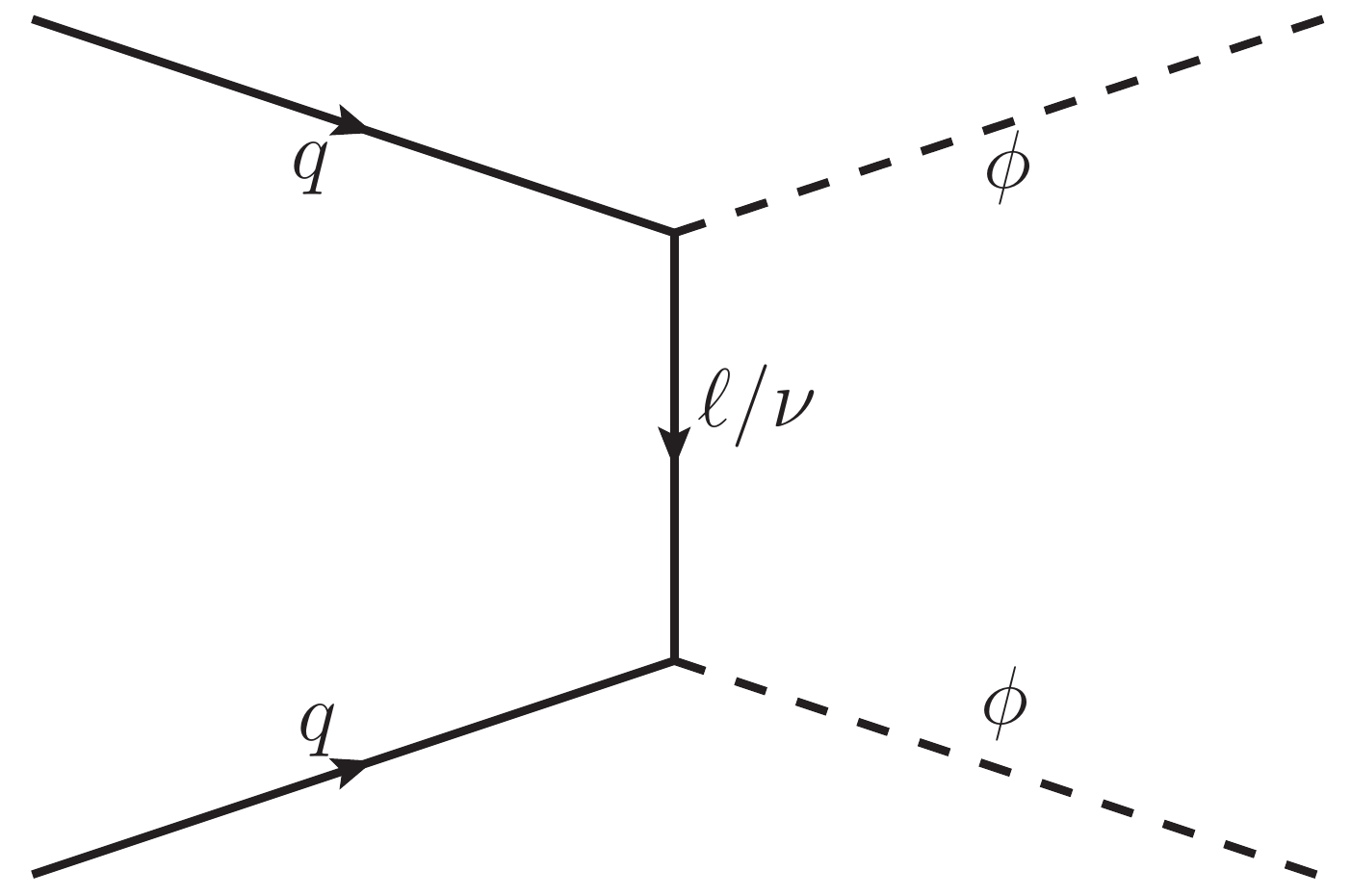}\label{fig:feynb}}\hfill
\subfloat[(c)]{\includegraphics[height=3cm,width=4cm]{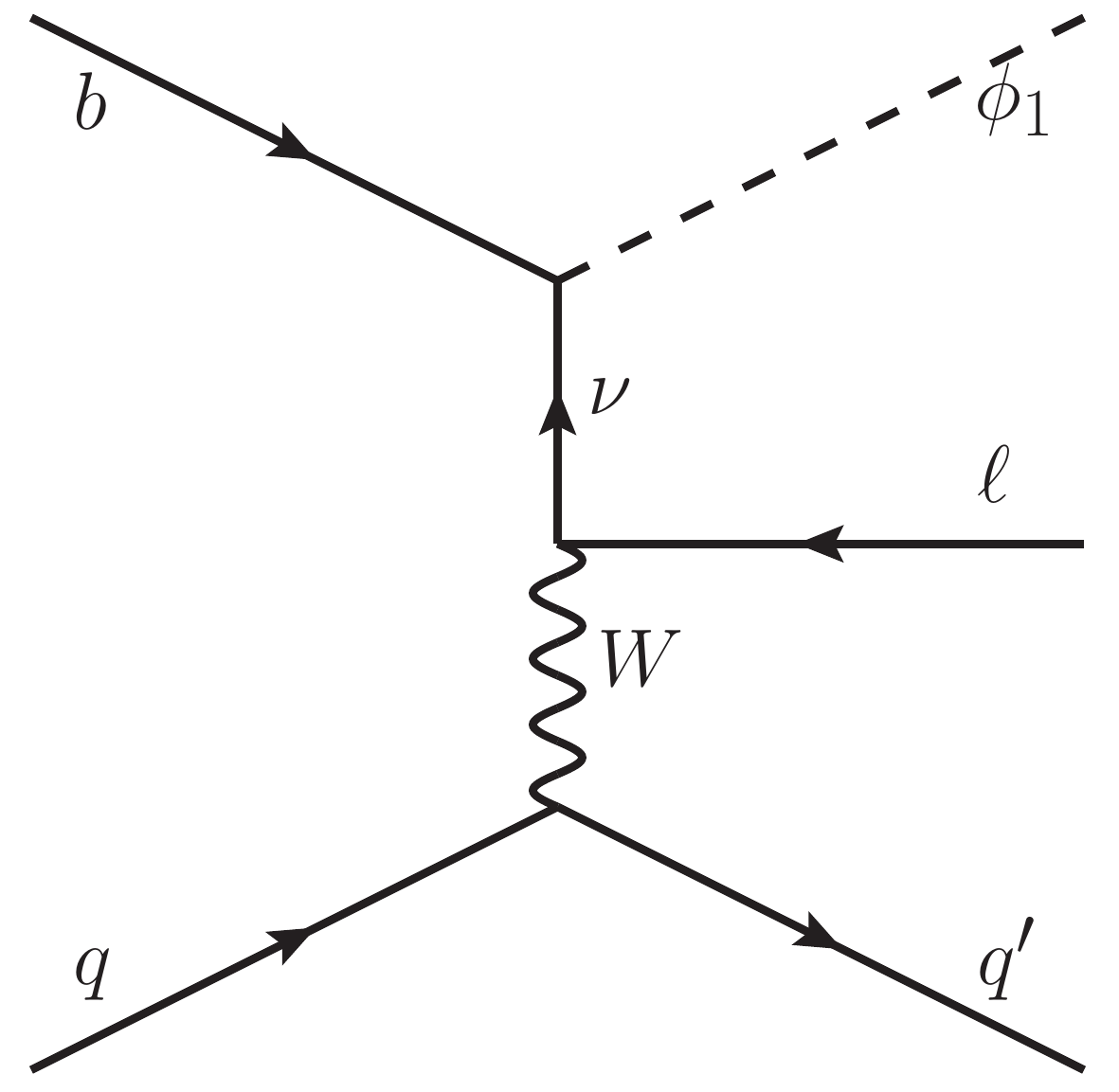}\label{fig:feync}}\hfill
\subfloat[(d)]{\includegraphics[height=3cm,width=4cm]{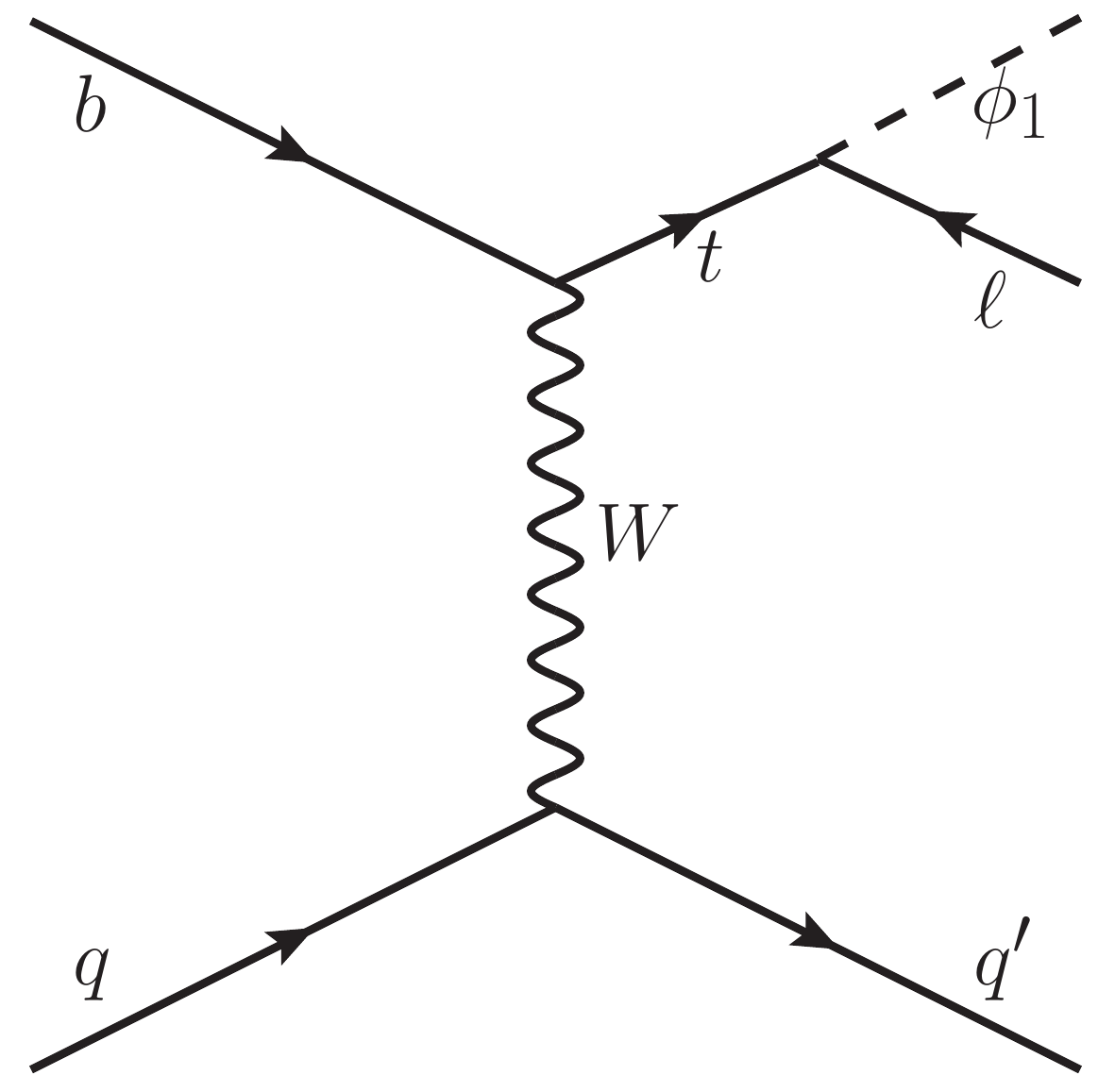}\label{fig:feynd}}
\caption{Sample Feynman diagrams for LQ production at the LHC. Diagrams (a) and (b) show pair production processes and (c) and (d) are examples of single productions.}\label{fig:prod}
\end{figure*}

%%%%%%%%%%%%%%%%%%%%%%%%%%%%%%%%%%%%%%%%%%%%%%%%%%%%%%%%%%%%%%%%%%%%%%%%%%%%%%%%%%%%%%%%%%%%%%%%%%%%%%%%%%%%%%%
\section{leptoquark models}
\label{sec:model}

\noindent
Electromagnetic charge conservation forces the LQs that
decay to a top quark and a charged lepton to have electromagnetic 
charge $\pm 1/3$ or $\pm 5/3$. From the classification of possible LQ states in 
Refs.~\cite{Buchmuller:1986zs,Dorsner:2016wpm}, we see that only $S_1$, $R_2$ and $S_3$ have
the desired decay modes, $\ell_q\to t\ell$ (where $\ell=\{e,\mu\}$). 
Below, we show these three types of LQs Lagrangians following the notations of Ref.~\cite{Dorsner:2016wpm}. To avoid proton decay constraints, we ignore the diquark operators.

\subsection{Existing Models}
\label{sec:existingmodels}
\noindent
$\blacksquare\quad$\underline{$S_1=(\overline{\mathbf{3}},\mathbf{1},1/3)$:}\\
For $S_1$, one can write the following two renormalizable operators invariant under the SM gauge group ($G_{\mathrm{SM}}$):
\begin{align}
\label{eq:LagS1}
\mathcal{L} \supset &~ y^{LL}_{1\,ij}\bar{Q}_{L}^{C\,i} S_{1} i\tau^2 L_{L}^{j}+y^{RR}_{1\,ij}\bar{u}_{R}^{C\,i} S_{1} e_{R}^{j}+\textrm{h.c.},
\end{align}
where $Q_L$ and $L_L$ are the SM left-handed quark and lepton doublets, respectively. The superscript $C$  denotes  
charge conjugation. The Pauli 
matrices are represented by $\tau^k$ with $k=\{1,~2,~3\}$. Here, the generation indices are denoted 
by $i,j=\{1,\ 2,\ 3\}$.  
This can be written explicitly as,
\begin{align}
\mathcal{L} \supset &-(y^{LL}_1 U)_{ij} \bar{d}_{L}^{C\,i} S_{1} \nu_{L}^{j}+(V^T y^{LL}_1)_{ij}\bar{u}_{L}^{C\,i} S_{1} e_{L}^{j}\nn\\&+y^{RR}_{1\,ij}\bar{u}_{R}^{C\,i} S_{1} e_{R}^{j}+\textrm{h.c.}.
\end{align}
where $U$ and $V$ represent the  Pontecorvo-Maki-Nakagawa-Sakata (PMNS) neutrino mixing matrix and the Cabibbo-Kobayashi-Maskawa (CKM) quark mixing matrix, respectively. Since the neutrino flavours cannot be distinguished at the LHC, we denote them by just $\nu$. Similarly, for LHC phenomenology in general, and in particular for our analysis, the small off-diagonal terms of the CKM matrix play negligible role. Hence, we assume a diagonal CKM matrix for simplicity. We identify the terms relevant for our analysis,
\begin{align}
\mathcal{L} \supset &\ y^{LL}_{1\ 3j} \left(-\bar{b}_{L}^C \nu_{L}+\bar{t}_{L}^{C}\ell_{L}^{j} \right)S_{1}
+y^{RR}_{1\,3j}\ \bar{t}_{R}^{C} \ell_{R}^{j} S_{1}+\textrm{h.c.},\label{eq:LagS1us}
\end{align}
where $j=\{1,\ 2\}$.
\bigskip

\noindent
$\blacksquare\quad$\underline{$S_3= (\overline{\mathbf{3}},\mathbf{3},1/3)$:}\\
There is only one type of $G_{\mathrm{SM}}$-invariant renormalizable operator one can write for $S_3$:
\begin{align}
\label{eq:LagS3}
\mathcal{L} \supset &~y^{LL}_{3\,ij}\bar{Q}_{L}^{C\,i,a} \epsilon^{ab} (\tau^k S^k_{3})^{bc} L_{L}^{j,c} + \textrm{h.c.},
\end{align}
Here, the $\mathrm{SU}(2)$ indices are denoted 
by $a,b,c=\{1,\ 2\}$.  Expanding this we get,
\begin{align}
\mathcal{L} \supset &-(y^{LL}_{3}U)_{ij}\bar{d}_{L}^{C\,i} S^{1/3}_{3} \nu_{L}^{j}-\sqrt{2} y^{LL}_{3\,ij}\bar{d}_{L}^{C\,i} S^{4/3}_{3} e_{L}^{j}\nonumber\\
&+\sqrt{2} (V^Ty^{LL}_{3}U)_{ij}\bar{u}_{L}^{C\,i} S^{-2/3}_{3} \nu_{L}^{j}\nn\\
&-(V^Ty^{LL}_{3})_{ij}\bar{u}_{L}^{C\,i} S^{1/3}_{3} e_{L}^{j}+\textrm{h.c.},
\end{align}
The relevant interaction terms can be written as,
\begin{align}
\mathcal{L} \supset & -y^{LL}_{3\ 3j}\left[\left(\bar{b}_{L}^{C} \nu_{L}+\bar{t}_{L}^{C} \ell_{L}^{j}\right) S^{1/3}+\sqrt{2}\left( \bar{b}_{L}^{C}  \ell_{L}^{j}S^{4/3}_{3}\right.\nn\right.\\
&-
\left.\left.\bar{t}_{L}^{C} \nu_{L}S^{-2/3}\right)\right]+\textrm{h.c.},\label{eq:LagS3us}
\end{align}
with $j=\{1,\ 2\}$.
\bigskip

\noindent
$\blacksquare\quad$\underline{$R_2=(\mathbf{3},\mathbf{2},7/6)$:}\\Similarly, for $R_2$ we have the following terms,
\begin{eqnarray}\label{eq:LagR2}
\mathcal{L} &\supset&-y^{RL}_{2\,ij}\bar{u}_{R}^{i} R_{2}^{a}\epsilon^{ab}L_{L}^{j,b}+y^{LR}_{2\,ji}\bar{e}_{R}^{j} R_{2}^{a\,*}Q_{L}^{i,a} +\textrm{h.c.},\ \nn
\end{eqnarray}
which, after expansion, can be written as,
\begin{eqnarray}
\nonumber
\mathcal{L} &\supset&  - y^{RL}_{2\ ij} \bar{u}^i_R e^j_L R_2^{5/3}+  (y^{RL}_2
  U)_{ij} \bar{u}^i_R 
  \nu^j_L R_2^{2/3}\\
 &&+(y^{LR}_2 V^\dagger)_{ji} \bar{e}^j_R 
  u^i_L R_2^{5/3\,*} +y^{LR}_{2\,ji} \bar{e}^j_R d^i_L R_2^{2/3\,*} + \textrm{~h.c.}.\;\;
\end{eqnarray}
We identify the terms relevant for us as,
\begin{eqnarray}
\mathcal{L} &\supset&  - y^{RL}_{2\ 3j}\ \bar{t}_R \ell^j_L R_2^{5/3}+  y^{RL}_{2\ 3j}\ \bar{t}_R 
  \nu_L R_2^{2/3}\nn\\
 &&+y^{LR}_{2\ j3}\ \bar{\ell}^j_R 
  t_L R_2^{5/3\,*} +y^{LR}_{2\ j3}\ \bar{\ell}^j_R b_L R_2^{2/3\,*} + \textrm{h.c.},\label{eq:main_R_2_a}
\end{eqnarray}
with $j=\{1,\ 2\}$.

%%%%%%%%%%%%%%%%%%%%%%%%%%%%%%%%%%%%%%%%%%%%%%%%%%%%%%%%%%%%%%%%%%%%%%%%%%%%%%%%%%%%%%%%%%%%%%%%%%%%%%%%%%%%%%%
\subsection{Simplified Models and Benchmark Scenarios}
\label{subsec:benchmark}
\begin{table*}
\begin{tabular}{|c|c|c|c|c|c|c|c|c|c|}\hline
&&\multicolumn{3}{c|}{Simplified model [Eqs.~\eqref{eq:simplelag1} -- \eqref{eq:simplelag3}]}&\multicolumn{2}{c|}{LQ models [Eqs.~\eqref{eq:LagS1us} -- \eqref{eq:main_R_2_a}]}&&\\\cline{3-7}
\begin{tabular}[c]{c}Benchmark \\ scenario\end{tabular} & \begin{tabular}[c]{c}Possible \\ charge(s)\end{tabular} & \begin{tabular}[c]{c}Type of  \\ LQ\end{tabular} & \begin{tabular}[c]{c}Non-zero \\ couplings\\ equal to $\lm$\end{tabular} & \begin{tabular}[c]{c}Lepton \\ chirality \\fraction\end{tabular} & \begin{tabular}[c]{c}Type of  \\ LQ\end{tabular} &\begin{tabular}[c]{c}Non-zero \\ coupling\\ equal to $\lm$\end{tabular} & \begin{tabular}[c]{c}Decay \\ mode(s)\end{tabular}& \begin{tabular}[c]{c}Branching \\ ratio(s)\end{tabular}\\\hline\hline
LCSS   	&$1/3$			& $\phi_1$     			&	$\lm_\ell=\lm_\n$         	& $\eta_L=1$, $\eta_R=0$		& $S_3^{1/3}$     					&$-y^{LL}_{3\ 3j}$		& \{$t\ell$, $b\n$\} 	& \{$50\%$, $50\%$\}  \\
LCOS   	&$1/3$			& $\phi_1$     			& $\lm_\ell=-\lm_\n$ 			& $\eta_L=1$, $\eta_R=0$ 		& $S_1$    					&$y_{1\ 3j}^{LL}$ 		&  \{$t\ell$, $b\n$\} & \{$50\%$, $50\%$\}         \\
RC       		&$\{1/3,5/3\}$		& $\{\phi_1,\phi_5\}$  & $\{\tilde\lm_\ell,\lm_\ell\}$		& $\eta_L=0$, $\eta_R=1$ 		& $\{S_1$, $R_2^{5/3}\}$ 	 		&$\{y_{1\ 3j}^{RR}$, $y^{LR}_{2\ j3}\}$		&  $t\ell$ & $100\%$ \\
LC       		&$5/3$			& $\phi_5$     			& $\tilde\lm_\ell$					& $\eta_L=1$, $\eta_R=0$		& $R_2^{5/3}$   					&$-y^{RL}_{2\ 3j}$		&  $t\ell$& $100\%$ \\ \hline   
\end{tabular}
\caption{Summary of the four benchmark scenarios considered. They are explained in Section~\ref{subsec:benchmark}.  }\label{tab:benchmark}
\end{table*}

\noindent
Following Ref.~\cite{Mandal:2015vfa}, we write a simplified phenomenological Lagrangian for the  models above,
\begin{eqnarray}
\mc{L} &\supset& \lm_{\ell}\lt(\sqrt{\eta_L}\bar{t}_L^C\ell_L
 + \sqrt{\eta_R}\bar{t}_R^C\ell_R\rt)\phi_1+\lm_\nu \bar{b}_L^C\nu_L\phi_1+\ \textrm{h.c.},\ \ \ \ \label{eq:simplelag1}
\\
\mc{L} &\supset& \tilde\lm_{\ell}\lt(\sqrt{\eta_L}\bar{t}_R\ell_L + \sqrt{\eta_R}\bar{t}_L\ell_R\rt)\phi_5+\textrm{h.c.}.\label{eq:simplelag3}
\end{eqnarray}
In this notation, a charge $1/3$ ($5/3$) scalar LQ is generically represented by $\phi_1$ ($\phi_5$). Here, $\eta_L$ and $\eta_R = (1 - \eta_L)$ are the
fractions of leptons coming from LQ decays that are left-handed and right-handed, respectively.
The simplified Lagrangian does not include any charge $2/3$ or $4/3$ LQ as such LQs would not couple  with just a top quark and a charged lepton simultaneously.

For our analysis, we consider four benchmark coupling scenarios.
\begin{enumerate}
    \item {\bf Left-handed Couplings with Same Sign (LCSS):} In this scenario, we set $\lm_\ell=\lm_\n=\lm$, $\tilde\lm_\ell=0$ and $\eta_R=0$, i.e., we have a $\phi_1$ LQ that couples to the left-handed leptons. As a result, it couples to both $t\ell$ and $b\nu$ pairs with equal strength and hence decays to either of the pairs with about $50\%$ BRs. In this scenario, the $\phi_1$ behaves like the charge $1/3$ component of $S_3$ (with $-y^{LL}_{3\ 3j}=\lm$).
    
    \item {\bf Left-handed Couplings with Opposite Sign (LCOS):} We set $\lm_\ell=-\lm_\n=\lm$, $\tilde\lm_\ell=0$ and $\eta_R=0$. In this scenario too a $\phi_1$ LQ couples with the left-handed leptons equally but with opposite signs. However, since it couples to both $t\m$ and $b\nu$ pairs with equal (absolute) strength, it still decays to either a $t\ell$ or a $b\nu$ pair with about $50\%$ BRs. In this scenario, it behaves like an $S_1$ with $y_{1\ 3j}^{LL}=\lm$ and $y_{1\ 3j}^{RR}=0$.
    
    \item {\bf Right-handed Coupling (RC):} In this scenario, the LQ has no weak charge and couples with only right handed leptons.  This scenario is common to both $\phi_1$ and $\phi_5$ as we do not use the charge of leptoquark in our analysis. Here, we set  $\tilde\lm_\ell=\lm_\ell=\lm$, $\lm_\n=0$ and $\eta_L=0$. It decays to a $t\ell$  pair with  $100\%$ BR. In this scenario, the LQ is either of $S_1$ type with $y_{1\ 3j}^{LL}=0$ and $y_{1\ 3j}^{RR}=\lm$ or it is $R_2^{5/3}$ with $y^{LR}_{2\ j3}=\lm$.
    
     \item {\bf Left-handed Coupling (LC):} In this scenario the LQ couples with only left handed charged leptons.  This scenario is exclusive to $\phi_5$. Here, we set $-y^{RL}_{2\ 3j}=\tilde\lm_\ell=\lm$, $\lm_\ell=\lm_\n=0$ and $\eta_R=0$. It decays to a $t\ell$  pair with  $100\%$ BR. 
\end{enumerate}
We have summarized these four scenarios in Table~\ref{tab:benchmark}.

%%%%%%%%%%%%%%%%%%%%%%%%%%%%%%%%%%%%%%%%%%%%%%%%%%%%%%%%%%%%%%%%%%%%%%%%%%%%%%%%%%%%%%%%%%%%%%%%%%%%%%%%%%%%%%%
\section{LHC Phenomenology \& Search Strategy}
\label{sec:pheno}

\noindent
We have used various publicly available packages for our analysis. We implement the Lagrangian of Eqs. \eqref{eq:simplelag1} and \eqref{eq:simplelag3} in {\tt FeynRules}~\cite{Alloul:2013bka} to create the UFO~\cite{Degrande:2011ua}
model files. Both the signal and the background events are generated in the event generator {\tt MadGraph5}~\cite{Alwall:2014hca} at the leading order (LO). The 
higher-order corrections are included by multiplying appropriate QCD $K$-factors wherever available. %~\cite{Mandal:2015lca}. 
We use NNPDF2.3LO~\cite{Ball:2012cx} parton distribution 
functions (PDFs) for event generation by setting default dynamical renormalization and factorization scales used in
{\tt MadGraph5}. Events are passed through {\tt Pythia6}~\cite{Sjostrand:2006za} to perform showering and 
hadronization and matched up to two additional jets using {\tt MLM} matching 
scheme~\cite{Mangano:2006rw,Hoche:2006ph} with virtuality-ordered Pythia showers to remove the double counting 
of the matrix element partons with parton showers. Detector effects are simulated using 
{\tt Delphes3}~\cite{deFavereau:2013fsa} with the default CMS card.
Fatjets are reconstructed using the {\tt FastJet}~\cite{Cacciari:2011ma} package by clustering {\tt Delphes} 
tower objects. We employ Cambridge-Achen~\cite{Dokshitzer:1997in} algorithm with radius parameter $R = 1.5$
for fatjet clustering. To reconstruct hadronic tops from fatjets, we use a popular top tagger, namely the {\tt HEPTopTagger}~\cite{Plehn:2010st}.

%%%%%%%%%%%%%%%%%%%%%%%%%%%%%%%%%%%%%%%%%%%%%%%%%%%%%%%%%%%%%%%%%%%%%%%%%%%%%%%%%%%%%%%%%%%%%%%%%%%%%%%%%%%%%%%
\subsection{Production at the LHC}

\begin{figure*}[]
\captionsetup[subfigure]{labelformat=empty}
\subfloat[\quad\quad\quad(a)]{\includegraphics[scale=0.66]{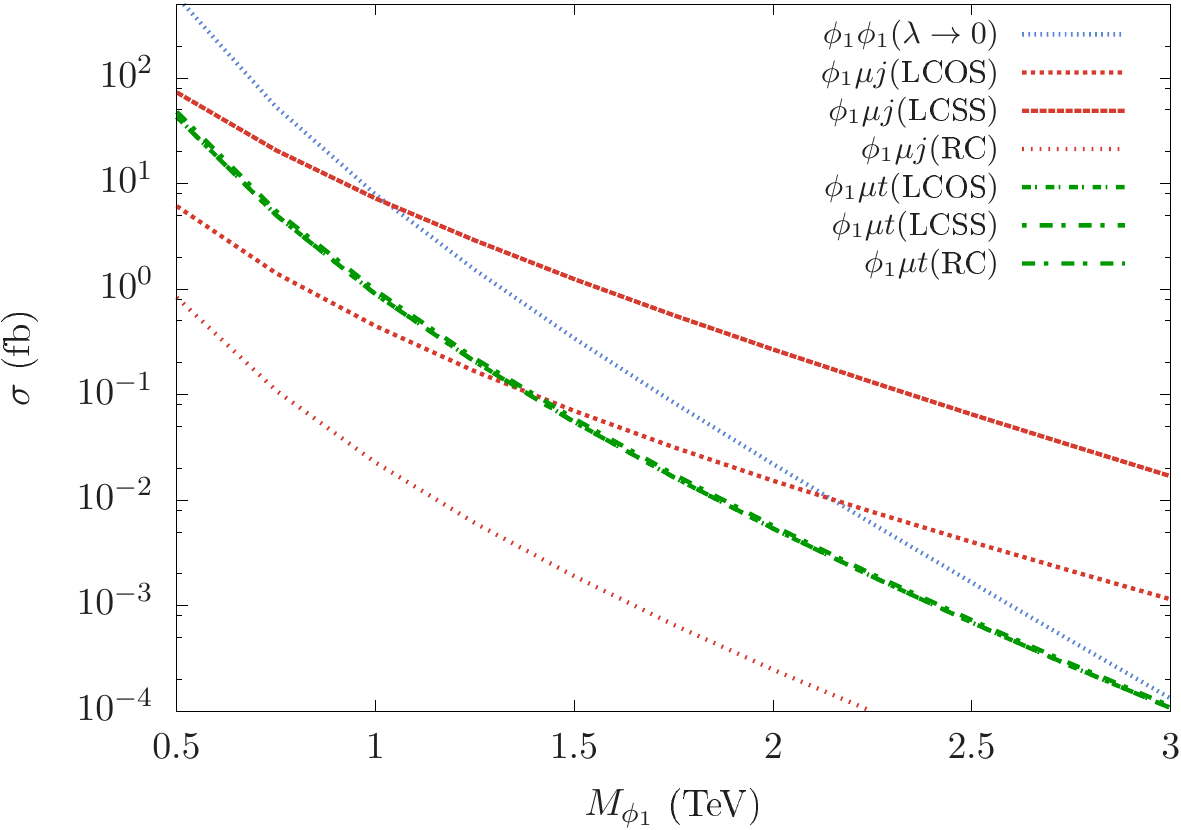}\label{fig:xseca}}\hfill{}
\subfloat[\quad\quad\quad(b)]{\includegraphics[scale=0.66]{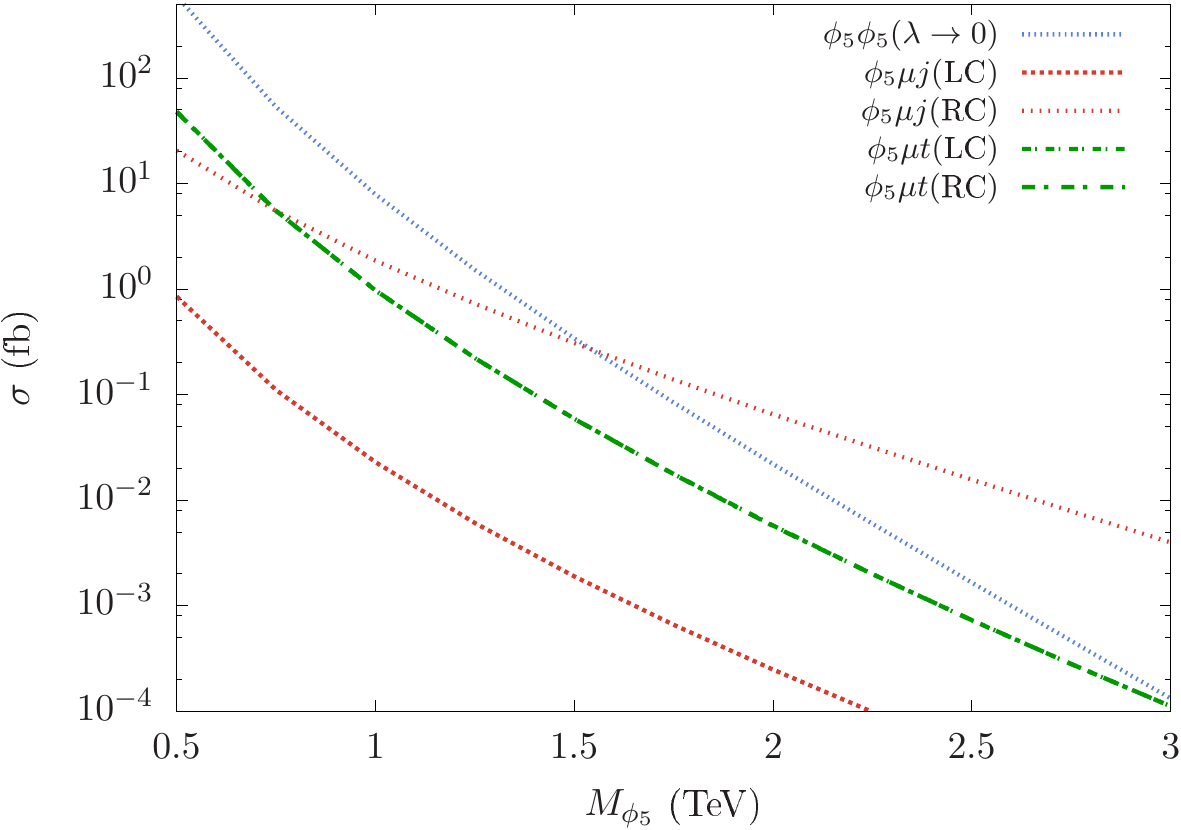}\label{fig:xsecb}}
\caption{The parton-level cross sections of different production channels of $\phi_1$ and $\phi_5$ at the $14$ TeV LHC as functions of $M_{\phi}$. We display the muon channel cross sections; the electron channel have similar cross sections. The single production cross sections are computed for a benchmark coupling $\lm=1$ (see Table~\ref{tab:benchmark}). The pair production cross sections include an NLO QCD $K$-factor of 1.3~\cite{Mandal:2015lca}. Here, the $j$ in the single production processes includes all the light jets as well as $b$-jets. Their cross sections are generated with a cut on the transverse momentum of the jet, $p^j_{\rm T} > 20$ GeV.}
\label{fig:xsec}
\end{figure*}

\noindent
As indicated in the \hyperref[sec:intro]{introduction} section, LQs are produced resonantly at the LHC through pair and single
production channels.  
The  pair production  is mostly model independent [depends only on the universal QCD
coupling, see e.g. Fig.~\ref{fig:feyna}] and proceeds through the $gg$ and $qq$ initiated processes. In the LCOS and the LCSS models, the process $bb\to \phi_1\phi_1$ through
the $t$-channel neutrino exchange is dependent on model coupling $\lm$ [see Fig.~\ref{fig:feynb}]. However, this contribution is small
in the total pair production cross section. The pair production process leads to the following final state,
\begin{align}
pp\to \phi\phi\to (t\ell)(t\ell)\label{eq:pair}
\end{align}
where a $\phi$ stands for either a $\phi_1$ or a $\phi_5$.
Single production channels, where a LQ is produced in association with a lepton and either a jet or a top-quark, are given as, 
\begin{eqnarray}
\left.\begin{array}{lcccl}
pp &\to &\phi t \ell &\to & (t\ell)t\ell \\
pp &\to & \phi\ell j &\to &  (t\ell)\ell j\label{eq:match_taunu}
\end{array}\right\}.
\end{eqnarray}

In Fig.~\ref{fig:xsec}, we show the parton level cross sections of different production processes of $\phi_1$ [Fig.~\ref{fig:xseca}] and $\phi_5$ [Fig.~\ref{fig:xsecb}]. The single productions are computed for  $\lm=1$.
We see that for $\phi_1$, the single production processes depend heavily on whether it is an $S_1$ with  LCOS/RC type couplings or an $S_3$ with LCSS coupling. In the LCSS scenario, the $pp\to\phi_1\ell j$ becomes the dominant process beyond $M_{\phi}\gtrsim 1$ TeV whereas in the LCOS scenario, it overtakes the pair production only for $M_{\phi}>2.2$ TeV. This difference happens since in the LCOS scenario, some single production diagrams [see e.g. Figs.~\ref{fig:feync} \& \ref{fig:feynd}] interfere destructively because of the opposite relative sign of the $\lm_\ell$ and $\lm_\n$ couplings, whereas in case of LCSS, they interfere constructively. In the RC scenario, $\phi_1$ does not couple to a $b$-quark or a left handed top quark (that can be produced from a $W$ boson and a $b$-quark interaction) and hence we do not expect $\sg(pp\to\phi_1\ell j)$ to be large. We see that $\sg(pp\to\phi_1\ell j)<\sg(pp\to\phi_1\phi_1)$ for $M_{\phi_1} < 3$ TeV in this scenario. For $\phi_5$, the cross section of $pp\to\phi_5\ell j$ processes in the LC scenario is smaller than that in the RC scenario, as $\phi_5$ couples exclusively to a right handed top quark in this case. 

It is clear from the cross section plots that for order one $\lm$, it is important to consider single productions while estimating the discovery prospects. Before we move on, we note that the cross section plots do not show the full picture, as one has to consider the branching ratios and the detector effects. In the LCOS and LCSS scenarios,  BR$(\phi\to t\ell) \sim 50\%$ whereas it is $100\%$ in the RC and LC scenarios. 

%%%%%%%%%%%%%%%%%%%%%%%%%%%%%%%%%%%%%%%%%%%%%%%%%%%%%%%%%%%%%%%%%%%%%%%%%%%%%%%%%%%%%%%%%%%%%%%%%%%%%%%%%%%%%%%
\subsection{Signal Topology}
\noindent
In our analysis, we only consider the hadronic decays of tops to reconstruct them in the final states. The characteristic of our signal is the presence of one or two boosted top quarks forming one/two top-like fatjets and two high-$p_{\rm T}$ leptons. From Eqs.~\eqref{eq:pair} and \eqref{eq:match_taunu}, we see that if we define our signal as events containing \emph{exactly two high-$p_{\rm T}$  same flavour opposite sign (SFOS) leptons and at least one hadronic top-like fatjet} in the final state then it would include both single and pair productions and enhance the sensitivity. 

There is some overlap between the pair and the single production processes. For example, at the parton level, a $t\ell\, t\ell$ final state can be produced from both the pair production process as well as the $pp\to \phi t\ell$ processes. Hence one has to be careful to avoid double counting while computing single productions~\cite{Mandal:2015vfa}. In our simulations we achieve this by ensuring that for any single production process both $\phi$ and $\phi^\dag$ are never on-shell simultaneously.

\begin{table}[t!]
\begin{center}
\begin{tabular}{|c|c|c|c|}
\hline
\multicolumn{2}{|c|}{Background } & $\sg$ & QCD\\ 
\multicolumn{2}{|c|}{processes}&(pb)&Order\\\hline\hline
$V + jets$  & $Z + jets$  &  $6.33 \times 10^4$& NNLO \\ \cline{2-4} 
 \cite{Catani:2009sm,Balossini:2009sa}                 & $W + jets$  & $1.95 \times 10^5$& NLO \\ \hline
$VV + jets$  & $WW + jets$  & 112.64& NLO\\ \cline{2-4} 
    \cite{Campbell:2011bn}               & $WZ + jets$  & 46.74 & NLO\\ \cline{2-4} 
                   & $ZZ + jets$  &  15.99 & NLO\\ \cline{1-4}
Single $t$  & $tW$  &  70.0 & N$^2$LO \\ \cline{2-4} 
\cite{Kidonakis:2015nna}                   & $tb$  & 218.0 & N$^2$LO\\ \cline{2-4} 
                   & $tj$  &  11.17 & N$^2$LO\\  \cline{1-4}
$tt$~\cite{Muselli:2015kba}  & $tt + jets$  & 835.61 & N$^3$LO\\ \cline{1-4}
\multirow{2}{*}{$ttV$~\cite{Kulesza:2018tqz}} & $ttZ$  &  1.045 &NLO+NNLL \\ \cline{2-4} 
                   & $ttW$  & 0.653& NLO+NNLL \\ \hline
\end{tabular}
\caption{Total cross sections for the background processes considered in our analyses.We use these cross sections to obtain the higher order $K$-factors. }
\label{tab:Backgrounds}
\end{center}
\end{table}

%%%%%%%%%%%%%%%%%%%%%%%%%%%%%%%%%%%%%%%%%%%%%%%%%%%%%%%%%%%%%%%%%%%%%%%%%%%%%%%%%%%%%%%%%%%%%%%%%%%%%%%%%%%%%%%
\subsection{The SM Backgrounds}
\noindent
The main SM background processes for this signal topology would be those which give two high-$p_{\rm T}$ leptons and
a top-like jet originating from an actual top quark or other jets (which can come from hadronic decays of the SM particles or from QCD jets). We see that the single $Z$ and $t t$ processes contribute dominantly.
Processes with large cross section containing single lepton can also act as a background if the
second lepton appear due to a jet misidentified as a lepton. However, due to very small misidentification rate, these
class of processes contribute negligibly to the total background. 

Although some backgrounds are seemingly huge (see Table~\ref{tab:Backgrounds}), events that would satisfy the final signal selection criterion
used in our analysis would actually come from a very specific kinematic region.
With this in mind, we generate the background processes with 
some strong generation level cuts, for better statistics and saving computation time,

\noindent
\underline{\emph{Generation level cuts:}}
\begin{enumerate}
\item $p_{\rm T}(\ell_1)> 250$ GeV,
\item the invariant mass of the lepton pair $M(\ell_1,\ell_2) > 115$ GeV (the $Z$-mass veto).
\end{enumerate}
Here, $\ell_1$ and $\ell_2$ denote the leptons with the highest and the second highest $p_{\rm T}$, respectively.
We discuss the different background processes in more detail below.

\begin{enumerate}
\item \underline{$V + jets$:}\\
Inclusive single vector boson ($V={Z,W}$) production processes in the SM have very large cross sections and
therefore, can act as potential backgrounds for our signal even if the cut efficiencies are extremely small.
There are two types of single vector boson processes that we consider as potential backgrounds. 

\begin{enumerate}

\item
\underline{$Z/\gamma^* + jets$:}
This background is generated by simulating the process, $pp\rightarrow  Z/\gamma^{*} + (0,1,2,3)$-$jets\to \ell\ell+jets$ 
matched up to three extra partons. Here, the two high-$p_{\rm T}$ leptons can arise from the leptonic decays of the
$Z$-boson and a top-like fatjet can originate from the QCD jets. Since the invariant mass of the two 
leptons peaks at $Z$-mass, this background is controlled by the $Z$-mass veto.

\item 
\underline{$W + jets$:}
This process also has huge cross section like the previous one, but it is a reducible background. 
We generate it by simulating the process, $pp\rightarrow  W + (0,1,2,3)$-$jets\to \ell\nu+jets$ matched up to three extra partons. 
Requirement of a top-like jet can be fulfilled if the QCD jets mimic as a top-jet. 
%This is the only single lepton category potential background we have considered. 
However, as we demand the second lepton also to have high $p_{\rm T}$ where the 
lepton misidentification efficiency becomes small, we found this background to be negligible.
\end{enumerate}

\item 
\underline{$VV + jets$:}\\
There are four types of diboson processes viz. $Z_\ell Z_h$, $W_h Z_\ell$, $W_\ell W_\ell$ and $Z_\ell H_h$ that can act as sources of two high-$p_{\rm T}$ leptons. The subscripts ``$\ell$'' and ``$h$'' represent leptonic and hadronic decay modes
respectively. In these cases, the required top-like jet can arise from the hadronic decay products of bosons or from
the QCD jets. Processes containing leptonically decaying $Z$ can be drastically reduced by applying $Z$ mass veto on
the invariant mass of the lepton pair. We do not consider the case where one lepton come from the vector boson decays
and the other appear due to jets misidentified as leptons. We generate matched event samples including up to two jets 
of these processes.  

\item 
\underline{$tt + jets$:}\\
The SM top pair production at the LHC can provide us two high-$p_{\rm T}$ leptons when both the tops decay leptonically.
Additionally, a top-like jet which arise from the QCD jets together with those two leptons can mimic our signal.
We find that, like the $Z$ background, this contribution is also significant in our case. A priori, the $t_\ell t_h$ process where one top decays leptonically and the other
hadronically can also contribute to the background. We generate this events by matching up to two additional jets.

\item
\underline{$ttV$:}\\
The SM processes with a top pair associated with a vector boson can act as backgrounds for our signal.  We consider
the following four cases viz. $t_\ell t_\ell Z_h$, $t_\ell t_\ell W_h$, $t_h t_h Z_\ell$, $t_h t_\ell W_\ell$ depending on 
the decays of tops and vector bosons. We generate these event samples without adding extra jets in the final state. 

\item
\underline{tW:}\\
The SM $pp\to t_\ell W_\ell$ process contains two leptons in the final state and contribute to the background for our 
signal. We generate this process using matching by adding up to two extra jets.

\end{enumerate}
In Table~\ref{tab:Backgrounds} we collect the total cross sections of the background processes computed at various orders of QCD available in the literature. From these we compute the $K$-factors and, as mentioned, scale the corresponding LO cross sections  in our analysis.

%%%%%%%%%%%%%%%%%%%%%%%%%%%%%%%%%%%%%%%%%%%%%%%%%%%%%%%%%%%%%%%%%%%%%%%%%%%%%%%%%%%%%%%%%%%%%%%%%%%%%%%%%%%%%%%
\subsection{Event selection}
\label{ssec:evntsel}
\noindent
We apply the following sets of cuts on the signal and background events sequentially.

\renewcommand\theenumi{\bfseries $\mc C_\arabic{enumi}$:}
\renewcommand\labelenumi{\theenumi}

\begin{figure}[!t]
\includegraphics[scale=0.66]{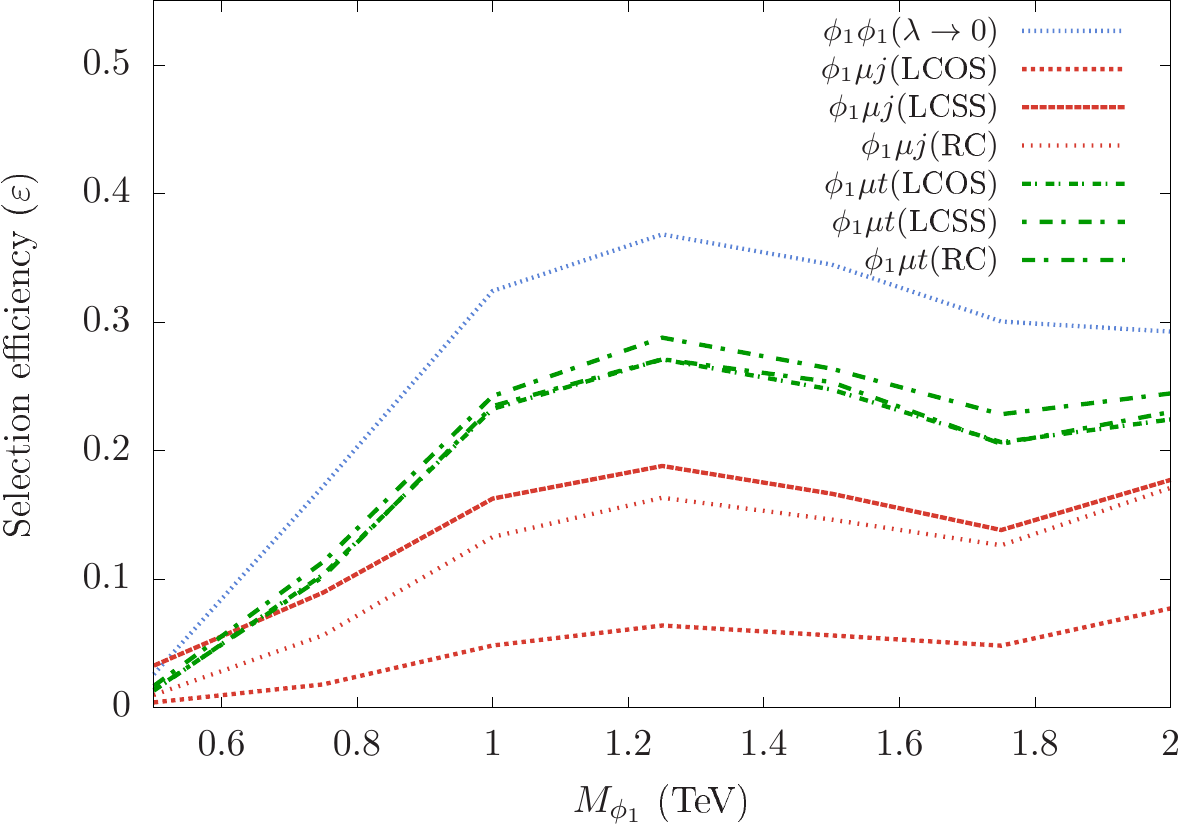}
\caption{Final  signal selection efficiencies ($\varepsilon$, see Eq.~\eqref{eq:eff}) for different coupling configurations in the $\mu$-channel. Since for $M_{\phi}\geq1.75$ TeV, the selection cuts do not change (see Section~\ref{ssec:evntsel}) we display the efficiencies only up to $M_{\phi}=2$ TeV. The $e$-channel efficiencies are very much similar to these.}\label{fig:SGCE}
\end{figure}

\begin{enumerate}\setcounter{enumi}{0}

\item 
{(a) At least one top-jet (obtained from HEPTopTagger) with $p_{\rm T}(t_h) > 135$ GeV.

(b) Two SFOS leptons with $p_{\rm T}(\ell_1)> 400$ GeV and $p_{\rm T}(\ell_2)> 200$ GeV 
and pseudorapidity $|\eta(\ell)|<2.5$. For electron we consider the barrel-endcap cut on $\eta$ between $1.37$ and $1.52$.

(c) Invariant mass of lepton pair $M(\ell_1,\ell_2) > 120$ GeV to avoid $Z$-peak.

(d) The missing energy $\slashed{E}_T < 200$ GeV.

}

\item The scalar sum of the transverse $p_{T}$ of all visible objects, $S_T>1.2\times {\rm Min}\lt( M_\phi, 1750\rt)$ GeV.

\item $M(\ell_1,t)~\mathrm{OR}~M(\ell_2,t) > 0.8\times {\rm Min}\lt(M_{\phi},1750\rt)$ GeV.
\end{enumerate}
In Fig.~\ref{fig:SGCE} we show the final signal selection efficiencies ($\varepsilon$) for different coupling hypotheses. We define $\varepsilon$ as,
\be
\varepsilon=\frac{\mbox{Number of events surviving } \mc C_1+\mc C_2+\mc C_3}{\mbox{Number of events generated}}.\label{eq:eff}
\ee
Since the $M_{\phi}$ dependent cuts (i.e., $\mc C_2$ and $\mc C_3$) get frozen beyond $M_{\phi}=1750$ GeV, we see the kink-like shapes at $1750$ GeV.

\begin{figure*}[!t]
\captionsetup[subfigure]{labelformat=empty}
\subfloat[\quad\quad\quad(a)]{\includegraphics[scale=0.63]{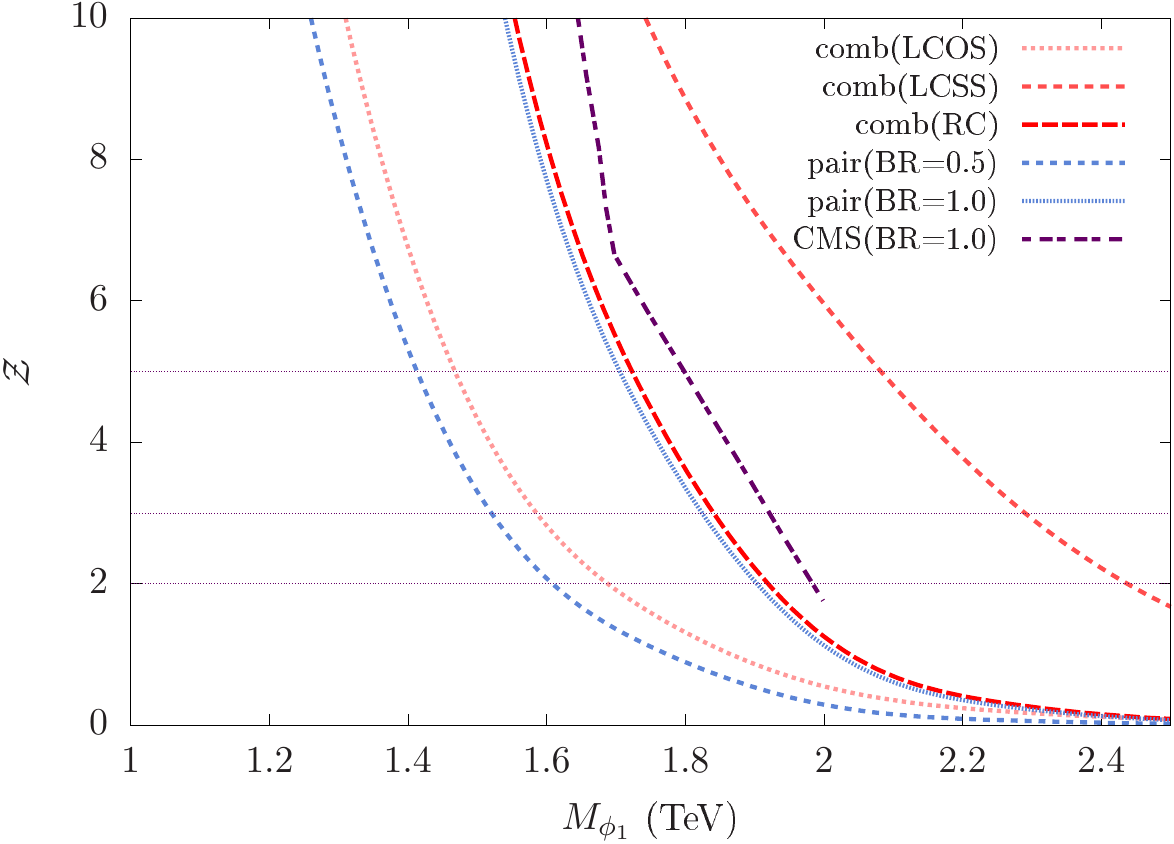}\label{fig:sigs1}}\hfill{}
\subfloat[\quad\quad\quad(b)]{\includegraphics[scale=0.63]{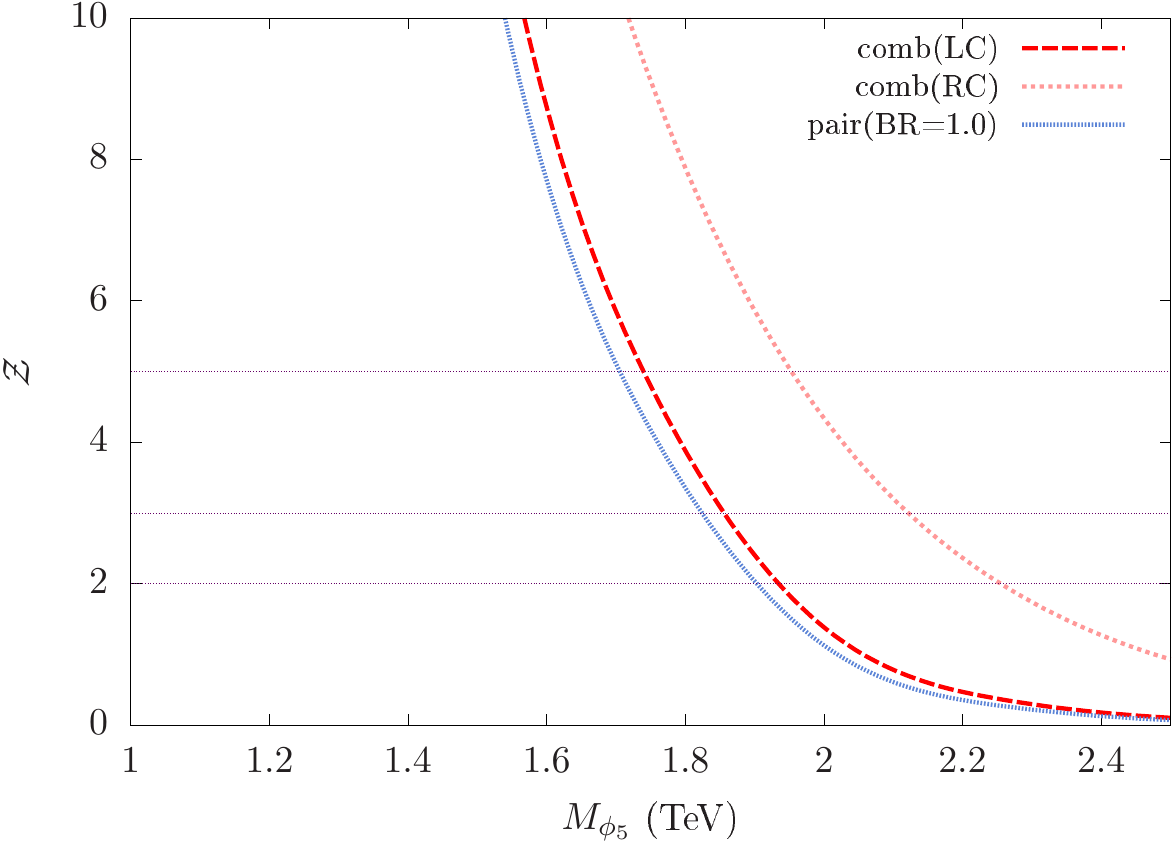}\label{fig:sigs5}}
\caption{Expected significance ($\mc Z$) for observing the  (a) $\phi_1$ and (b) $\phi_2$ signals over the SM backgrounds as functions of their masses for 3 ab$^{-1}$ of integrated luminosity at the 14TeV LHC for different coupling scenarios in the muon mode. The electron mode numbers can be seen from Table~\ref{tab:sig}. We use the combined pair and single productions for the signals in the LCOS, LCSS, RC and LC scenarios. For comparison, we also show the pair production  only significance for 50\% and 100\% BRs in the $\phi\to t\m$ decay mode and the CMS statistical-uncertainty-only estimation for discovering $\phi_1$~\cite{CMS:2018yke}. We have set $\lm=1$ while computing the combined signals. Our estimations are obtained using the event selection cuts defined in Section~\ref{ssec:evntsel}, i.e., only events with at least one hadronically decaying boosted top and two high-$p_T$ opposite sign electrons are considered.}
\label{fig:sig}
\end{figure*}

\begin{table*}[!t]
\begin{tabular}{|p{0.5cm} || c c c | c c | c c | c || c c c | c c | c c | c|}
\hline
 \multirow{3}{*}{~\rotatebox[origin=c]{90}{Significance $\mc Z$ }} & \multicolumn{16}{c|}{Limit on $M_\phi$ (TeV)}\\ 
	& \multicolumn{8}{c}{The $\mu$ channel}	& \multicolumn{8}{c|}{The $e$ channel} \\ \cline{2-17} 
                         & \multicolumn{5}{c|}{$\phi_1$} & \multicolumn{3}{c||}{$\phi_5$} & \multicolumn{5}{c|}{$\phi_1$} & \multicolumn{3}{c|}{$\phi_5$} \\ 
\cline{2-17}
&\multicolumn{3}{c|}{Combined}&\multicolumn{2}{c|}{Pair}&\multicolumn{2}{c|}{Combined}&Pair&\multicolumn{3}{c|}{Combined}&\multicolumn{2}{c|}{Pair}&\multicolumn{2}{c|}{Combined}&Pair\\
\cline{2-17}
&LCOS&LCSS&RC&BR=$0.5$&BR=$1.0$&LC&RC&BR=$1.0$&LCOS&LCSS&RC&BR=$0.5$&BR=$1.0$&LC&RC&BR=$1.0$\\
   \hline\hline
~~5 & 1.47 & 2.08 & 1.73 & 1.42 & 1.71 & 1.74 & 1.96 & 1.71 & 1.45 & 2.11 & 1.72 & 1.39 & 1.70 & 1.74 & 1.97 & 1.70 \\ \hline
~~3 & 1.59 & 2.29 & 1.84 & 1.52 & 1.83 & 1.86 & 2.12 & 1.83 & 1.58 & 2.33 & 1.84 & 1.52 & 1.83 & 1.86 & 2.16 & 1.83 \\ \hline
~~2 & 1.69 & 2.44 & 1.92 & 1.61 & 1.90 & 1.94 & 2.25 & 1.90 & 1.69 & 2.50 & 1.93 & 1.62 & 1.91 & 1.95 & 2.30 & 1.91 \\ \hline
\end{tabular}
\caption{The mass limits corresponding to $5\sg$ (discovery), $3\sg$ and $2\sg$ (exclusion) significances ($\mc Z$) for observing the  (a) $\phi_1$ and (b) $\phi_2$ signals over the SM backgrounds for 3 ab$^{-1}$ integrated luminosity at the 14TeV LHC with combined and pair only signals. The $\mu$-channel numbers can also be seen from Fig.~\ref{fig:sig}.}
\label{tab:sig}
\end{table*}

%%%%%%%%%%%%%%%%%%%%%%%%%%%%%%%%%%%%%%%%%%%%%%%%%%%%%%%%%%%%%%%%%%%%%%%%%%%%%%%%%%%%%%%%%%%%%%%%%%%%%%%%%%%%%%%
\section{Discovery potential}\label{sec:dispot}

\noindent
With the number of signal ($N_S$) and background ($N_B$) events surviving the selection cuts defined in Section~\ref{ssec:evntsel}, we estimate the expected significance ($\mc{Z}$) using the 
following formula:
\begin{align}
\mc{Z} =\sqrt{2\lt(N_S+N_B\rt)\ln\lt(\frac{N_S+N_B}{N_B}\rt)-2N_S}\, .
\end{align}
In Fig.~\ref{fig:sig}, we show the expected  significances for observing the $\phi_1$ and $\phi_2$ signals in the benchmark coupling scenarios (Section~\ref{subsec:benchmark}) over the SM backgrounds in the muon mode as functions of their masses for  3 ab$^{-1}$ of integrated luminosity at the 14 TeV LHC. 
As explained, we have used the combined signal (i.e.
pair and single production events together) to estimate the significances in the LCOS, LCSS, RC and LC scenarios with $\lm=1$. For the LCOS and LCSS scenarios, the BR of 
LQ to $te$ mode is 50\% whereas for the RC and LC scenarios it is 100\%. For comparison, we also show the expected significance for only the pair production 
(i.e. $\lm\to 0$) with 50\% and 100\% BR cases. In Table~\ref{tab:sig} we explicitly show the mass values corresponding to $5\sg$ (discovery), $3\sg$ and 
$2\sg$ (exclusion) significances for different coupling hypotheses in both $e$ and $\mu$ channels.

\begin{figure*}[!t]
\captionsetup[subfigure]{labelformat=empty}
\subfloat[\quad\quad\quad(a)]{\includegraphics[scale=0.63]{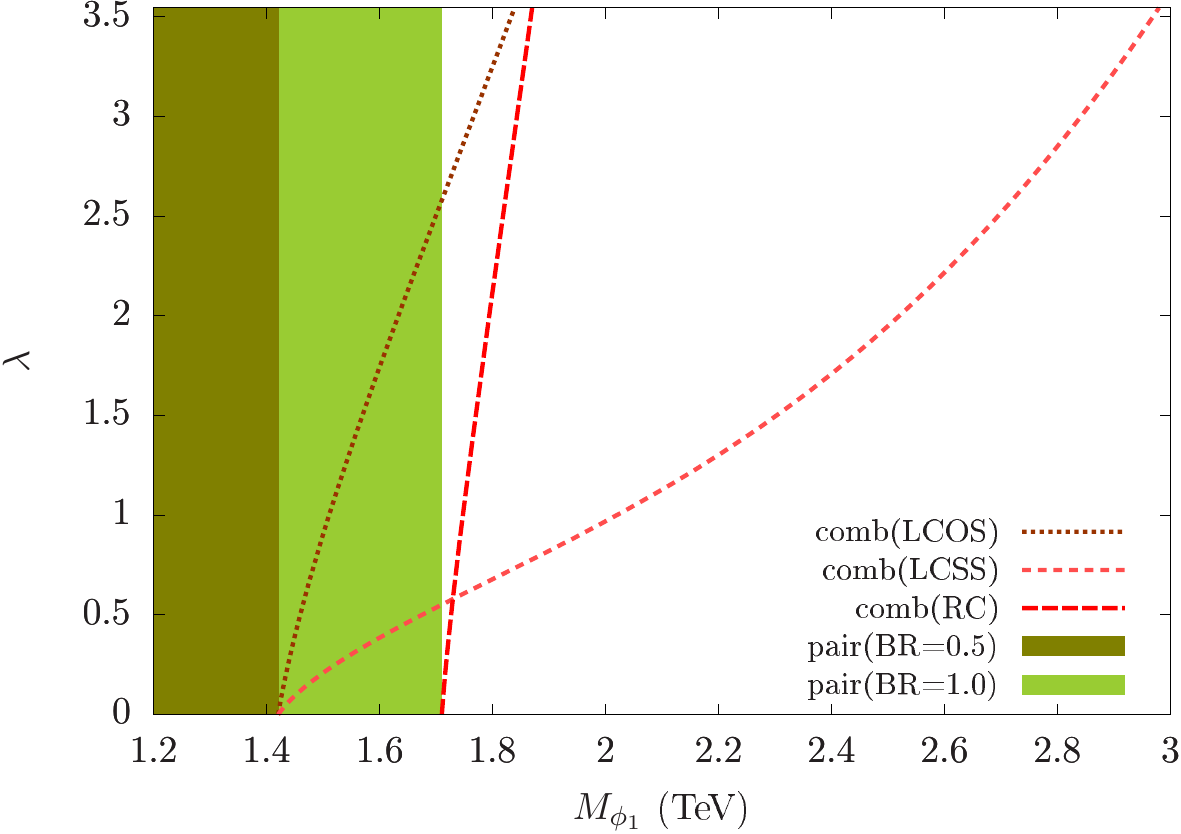}}\hfill{}
\subfloat[\quad\quad\quad(b)]{\includegraphics[scale=0.63]{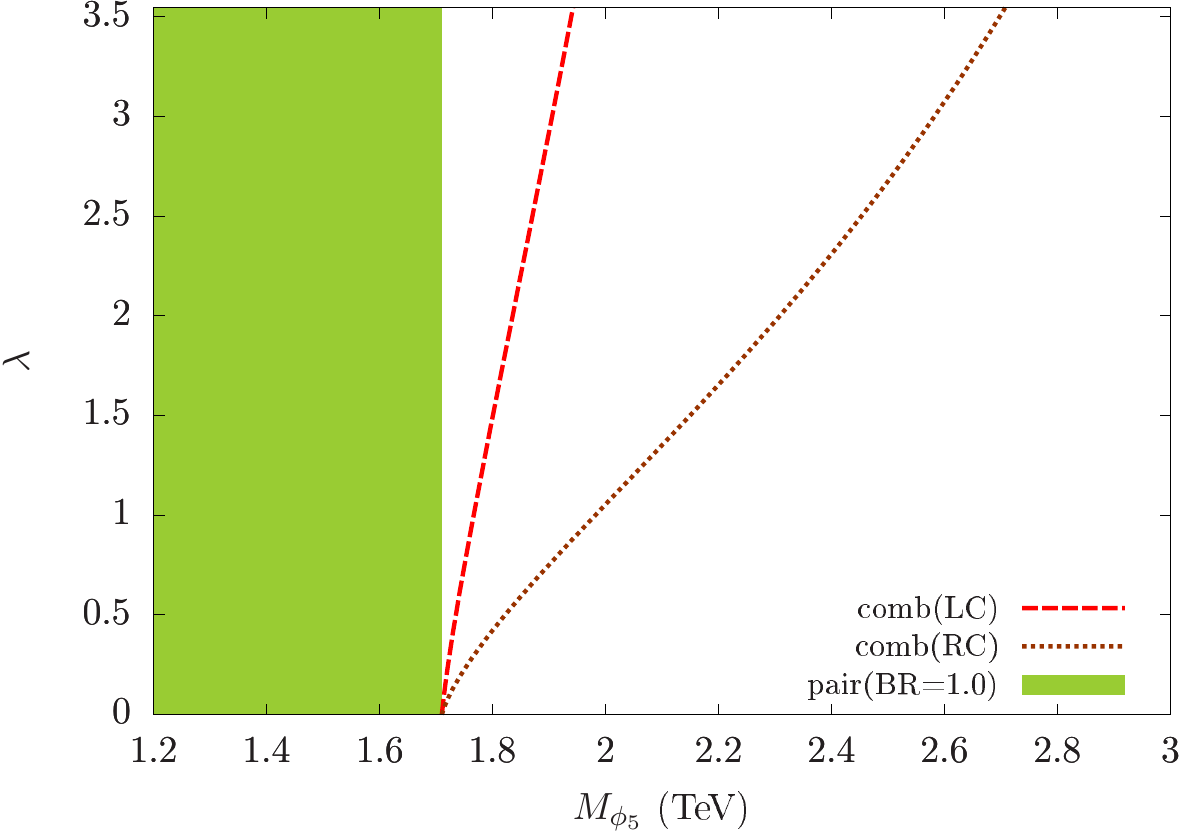}}
\caption{The $5\sg$ discovery reaches in the $\lm-M_\phi$ planes -- (a) for a charge $1/3$ scalar LQ and (b) for a charge $5/3$ scalar LQ. These plots show the lowest $\lm$ needed to observe  $\phi_1$ and $\phi_5$ signals with $5\sg$ significance for a range of $M_\phi$ with 3 ab$^{-1}$ of integrated luminosity. The pair production only regions for 50\% and 100\% BRs in the $\phi\to t\m$ decay mode are shown with shades of green. Since the pair production is insensitive to $\lm$, a small coupling is sufficient to attain $5\sg$ significance within the green regions.}
\label{fig:lamz5}
\end{figure*}

\begin{figure*}[!t]
\captionsetup[subfigure]{labelformat=empty}
\subfloat[\quad\quad\quad(a)]{\includegraphics[scale=0.63]{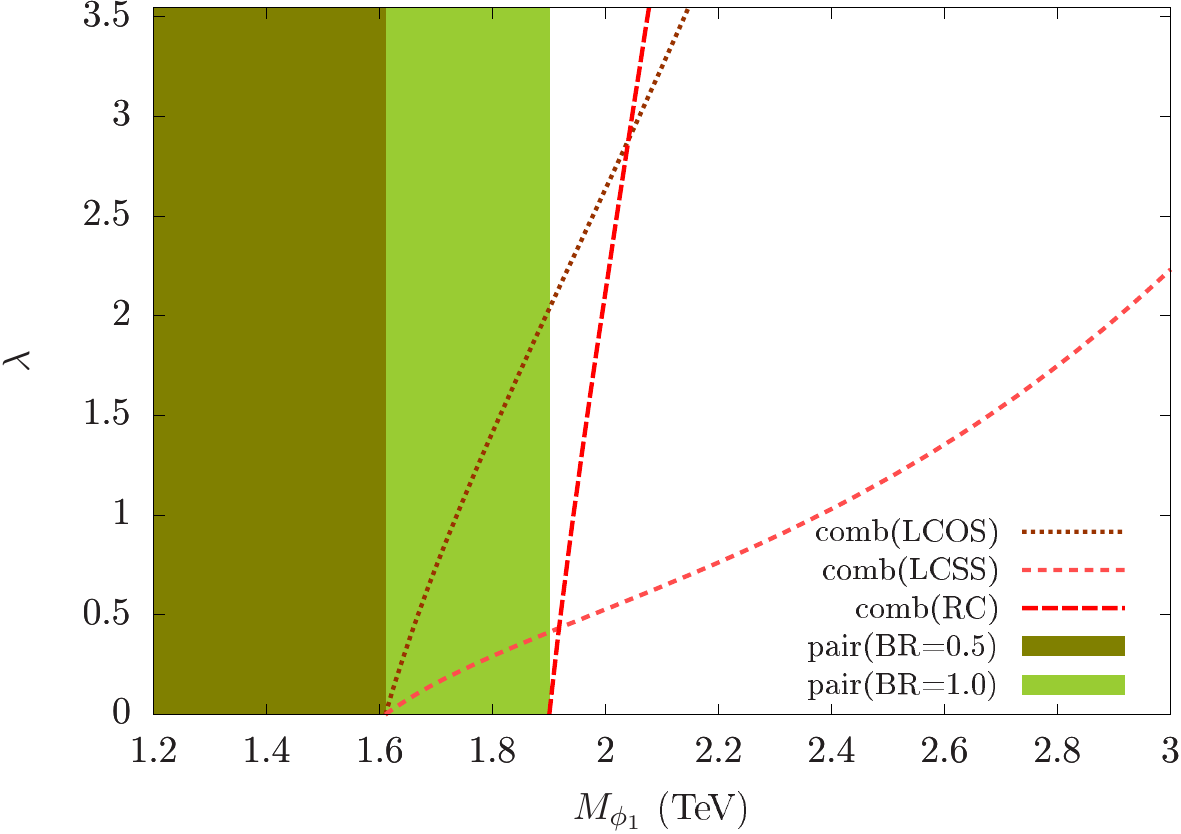}}\hfill{}
\subfloat[\quad\quad\quad(b)]{\includegraphics[scale=0.63]{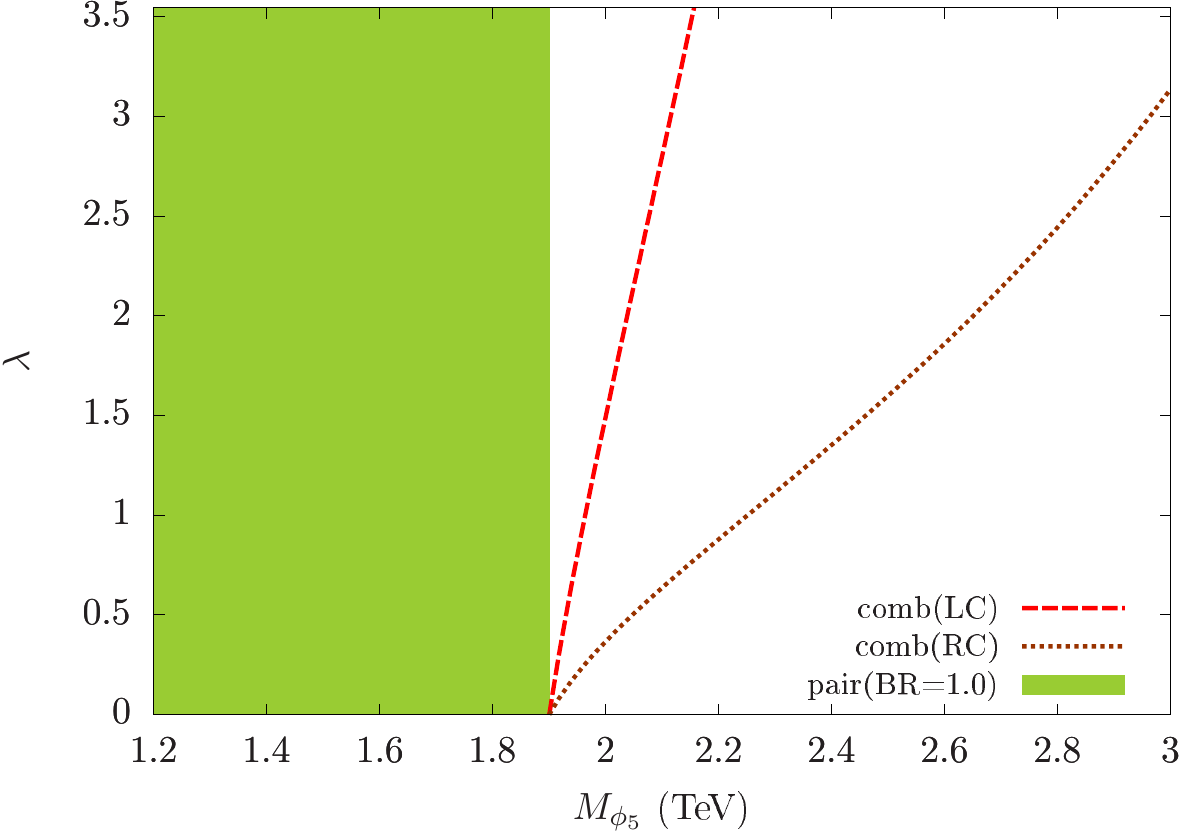}}
\caption{The $2\sg$ exclusion limits in the $\lm-M_\phi$ planes -- (a) for a charge $1/3$ scalar LQ and (b) for a charge $5/3$ scalar LQ. These plots show the lowest $\lm$ that can be excluded by the HL-LHC  with 3 ab$^{-1}$ of integrated luminosity. The pair production only regions for 50\% and 100\% BRs in the $\phi\to t\m$ decay mode are shown with green shades.}
\label{fig:lamz2}
\end{figure*}

As already mentioned, the CMS collaboration has projected the expected significance for scalar LQs decaying into $t\mu$ pairs in the pair production channel at the $14$ TeV HL-LHC~\cite{CMS:2018yke}. There, with $100$\% BR in the $t\mu$ mode, the $5\sg$ discovery reach goes to about $1.8$ TeV (considering statistical uncertainty only). Our estimate is quite close, $\sim1.7$ TeV if we consider only pair production with 100\% BR in the $\phi_1\to t\m$ decay
mode (see Table.~\ref{tab:sig}). This reach can decrease to $1.4$ TeV if the BR falls by $50$\%. However, if we include single productions, the $5\sg$ reach goes up to $2.1$ TeV in the LCSS scenario (where the $\phi_1$ behaves like the charge $1/3$ component of $S_3$). This drastic enhancement of $700$ GeV in the discovery reach happens because of the (relatively) large $pp\to \phi_1\ell j$ cross section in the high mass region leading to a substantial number of events surviving the applied selection cuts. However, in the LCOS scenario where a $\phi_1$ behaves like an $S_1$, this increment is minor, just about $50$ GeV, as destructive interference reduces the single production cross sections.

In the RC scenario, the total single production cross section of $\phi_1$ is small compared to the pair production one. Hence, the discovery reach is almost identical to that in the pair production only case. A similar situation is observed in the LC scenario for $\phi_5$. As explained in Section~\ref{sec:pheno}, in both the RC scenario for $\phi_1$ and the LC scenario for $\phi_5$, leptoquarks couple to the right-handed tops. As a result, single productions in these cases have small cross sections as right-handed tops can  couple to
the charged current only via chirality flipping. 

For any $M_\phi$ our signal cross section depends on $\lm$ as,
\be
\sg_{\rm signal} \approx \sg_{\rm pair}(M_\phi) + \lm^2 \sg_{\rm single}(\lm=1,M_\phi), 
\ee
i.e., for any $M_\phi$ if $\lm$ increases the signal increases. Using this relation one can recast the plots in Fig.~\ref{fig:sig} in the $\lm-M_\phi$ plane, as we have done in Fig.~\ref{fig:lamz5}. These plots show the lowest $\lm$ needed to observe  $\phi_1$ and $\phi_5$ signals with $5\sg$ significance for a range of $M_\phi$ with 3 ab$^{-1}$ of integrated luminosity. For all the points below a curve, the expected significance would be less than $5\sg$. In Fig.~\ref{fig:lamz2} we show the corresponding plots for $2\sg$ significance. In other words, these plot give us the lowest couplings that can be excluded at the HL-LHC.

%%%%%%%%%%%%%%%%%%%%%%%%%%%%%%%%%%%%%%%%%%%%%%%%%%%%%%%%%%%%%%%%%%%%%%%%%%%%%%%%%%%%%%%%%%%%%%%%%%%%%%%%%%%%%%%
\section{Summary and conclusions}\label{sec:End}

\noindent
In this paper, we have studied the HL-LHC reach for discovering scalar LQs that
decay to a top quark and a charged lepton. In particular, we have focused on
charge 1/3 ($\phi_1$) and 5/3 ($\phi_5$) scalar LQs that produce a resonance system with hadronically decaying boosted top quark and an electron or a muon. 
According to the classification given in Refs.~\cite{Buchmuller:1986zs,Dorsner:2016wpm}, only
 $S_1$, $S_3$ (charge 1/3 component of the triplet) and $R_2$ (charge 5/3 component of the doublet) scalar LQs can produce the specific signatures we consider. We have also introduced
a simplified Lagrangian for $\phi=\{\phi_{1},\phi_5\}$ suitable for bottom-up searches. We have shown how these simplified models connect to the actual models for different
coupling configurations. 

LQs can be produced in pairs or singly ($pp\to\phi \ell j,\phi \ell t$) at the LHC. When a LQ couples mostly with 
the third generation quarks, usually the pair production channels are considered for their discovery assuming the single
productions to be suppressed because of small $b$-PDF. Interestingly, we find that for order one couplings, cross section of
some single production channels can be larger than the pair production cross section. Hence, it is natural to expect that 
the inclusion of these single production channels would, therefore, increase their discovery prospects beyond the pair
production searches. However, this 
depends on the underlying model; 
the $pp\to\phi_1 \ell j$ single production cross sections can differ drastically depending on the representation in which 
$\phi_1$ belongs to.
In particular,
for the charge $1/3$ component of $S_3$ (the LCSS scenario, see Section~\ref{subsec:benchmark}), $pp\to\phi \ell j$ has larger cross section than the pair production in the heavier mass region. This also happens for $R_2^{5/3}$ (with $y^{LR}$ type coupling
 the RC scenario). 
However, for $S_1$ (the LCOS scenario), the single productions have smaller cross sections because of  destructive 
interference between certain diagrams [which is caused by the opposite signs of the left handed charged lepton and neutrino couplings, see Eq.~\eqref{eq:simplelag1}].

We have proposed a selection criterion that would retain events from both pair and single production processes so that the 
search becomes a combined one with increased reach. Our signal topology is defined by at least one hadronically decaying
boosted top and two opposite sign same flavour leptons. 
With this, we have found that the $5\sigma$ discovery reach for $\phi_1$ in LCSS scenario with $\lm=1$ is about 2.1 TeV at the 14 TeV LHC with 3 ab$^{-1}$ integrated
luminosity. In the LCSS scenario, the BR $\phi\to t\ell$ mode is 50\% and the reach for the pair production is only about 1.4 TeV. This significant improvement is due to constructive 
interference among certain single production diagrams. This increases the $pp\to \phi \ell j$ cross section about one order
in magnitude compared to the LCOS case where destructive interference  makes single production less important.
Finally we note that the enhancements of discovery reach due to the single production channels would increase further if the new couplings are more than one as the single production cross sections scale as square of the coupling involved. 

%%%%%%%%%%%%%%%%%%%%%%%%%%%%%%%%%%%%%%%%%%%%%%%%%%%%%%%%%%%%%%%%%%%%%%%%%%%%%%%%%%%%%%%%%%%%%%%%%%%%%%%%%%%%%%%
\acknowledgments 
\noindent 
T.M. is grateful to the Royal
Society of Arts and Sciences of Uppsala for financial support as a guest researcher at Uppsala University during the initial stage of this project. S.M. acknowledges support from the Science and
Engineering Research Board (SERB), DST, India under grant number ECR/2017/000517. We thank R. Arvind Bhaskar for reading and commenting on the manuscript.

%%%%%%%%%%%%%%%%%%%%%%%%%%%%%%%%%%%%%%%%%%%%%%%%%%%%%%%%%%%%%%%%%%%%%%%%%%%%%%%%%%%%%%%%%%%%%%%%%%%%%%%%%%%%%%%
\bibliography{reference}{}
\bibliographystyle{JHEPCust}

\end{document}